\documentclass[intlimits,twoside,a4paper]{article}

\usepackage{amsmath,amssymb}
\usepackage{graphicx}
\usepackage{wrapfig}

\usepackage[T2A]{fontenc}
\usepackage[cp1251]{inputenc}

\usepackage[eqsecnum]{cmpj2}

\issue{2012}{15}{2}{23607}

\doinumber{10.5488/CMP.15.23607}

\author[M.~Holovko, T.~Patsahan, W.~Dong]{M.~Holovko\refaddr{label1}, T.~Patsahan\refaddr{label1}, W.~Dong\refaddr{label2}}

\authorcopyright{M.~Holovko, T.~Patsahan, W.~Dong, 2012}

\addresses{
\addr{label1} Institute for Condensed Matter Physics of the
National Academy of Sciences of Ukraine, \\1 Svientsitskii Str., 79011 Lviv, Ukraine
\addr{label2} \'Ecole Normale Sup\'erieure de Lyon, Laboratoire de Chimie, UMR 5182 CNRS, \\
46 All\'ee d'Italie, 69364 Lyon, Cedex 07, France
}

\date{Received April 2, 2012, in final form May 9, 2012}

\title[A hard rod fluid in a disordered porous medium]%
{One-dimensional hard rod fluid in a disordered porous medium: scaled particle theory
}
\begin{document}

\maketitle

\begin{abstract}
The scaled particle theory is applied to a description of thermodynamic properties of
one-dimensional hard rod fluid in disordered porous media. To this end, we
extended the SPT2 approach, which had been developed previously. Analytical expressions are obtained for the chemical potential and pressure of a hard-rod
fluid in hard rod and overlapping hard rod matrices. A series of
new approximations for SPT2 are proposed.
It is shown that apart from two well known porosities such as geometrical
porosity and specific probe particle porosity, a new type of porosity defined by
the maximum value of packing fraction of fluid particles in porous medium
should be taken into account. The grand canonical Monte-Carlo simulations are
performed to verify the accuracy of the SPT2 approach in combination with the
new approximations. It is observed that the theoretical description proposed
in this study essentially improves the results up to the highest values of
fluid densities.
\pacs 61.20.Gy, 61.43.Gt
\end{abstract}

\section{Introduction}

It is a great pleasure for us to dedicate this paper to our good friend and
colleague Orest Pizio -- an excellent scientist in the study of fluids on
complex surfaces. A typical example of such complex systems is a fluid
confined in a disordered porous medium. Much theoretical effort has been
devoted to the study of fluids in porous materials for the last two decades
starting with a pioneering work by Madden and Glandt  \cite{Madd1}. In this
paper a simple model for a fluid adsorption in a disordered porous medium was
proposed. Within this model, a porous medium is presented as a matrix
composed of quenched configurations of model fluids.
Using the replica Ornstein-Zernike (ROZ) integral equations
  \cite{Giv2}, the statistical-mechanics approach of liquid state was extended
to a description of different fluids confined in disordered porous matrices
  \cite{Ros3,Pizio4}. An important contribution into the development of this approach
and its application has been made by O.Pizio with co-authors
  \cite{Pizio4,Trokh5,Trokh6,Pizio7,Hribar8,Kov9,Pad10,Hribar11,Malo12,Kov12}.
Despite a comprehensive study of a fluid in disordered matrices, no
analytical result has been obtained using the integral equation approach even
for a simple model like a hard sphere fluid in a hard sphere matrix. The main
complication in a description of such a model appears due to the absence of
a direct interaction between particles from different replicated copies of a
fluid. As a result, the description of this model is equivalent to the study of
a hard sphere mixture with strongly non-additive diameters, for which it is
very difficult to find a correct analytical result within the integral
equation approach.

The first rather accurate analytical results for a hard sphere fluid in hard
sphere (HS) and overlapping hard sphere (OHS) matrices were obtained quite
recently  \cite{Hol13,Hol13_2,Chen15,Pat16} by extending the scaled particle theory
(SPT)  \cite{Reiss17,Reiss18,Lebow19} to a HS fluid confined in a disordered
porous media. The SPT approach is based on a combination of the exact
treatment of a point scaled particle in a HS fluid and the thermodynamical
consideration of a finite size scaled particle. Exact results for a point
scaled particle in a HS fluid in disordered porous media were obtained in
  \cite{Hol13}. However, the proposed approach referred to as SPT1 has got  a subtle
inconsistency appearing when the size of matrix particles is essentially
larger than the size of fluid particles. This inconsistency was eliminated and
thus a new improved approach was developed referred to as SPT2
  \cite{Pat16}. The expressions obtained in SPT2 include two types of
porosities. One of them is defined by the pure geometry of porous medium
(geometrical porosity $\phi_{0}$) and the second one is defined by the
chemical potential of a fluid in the limit of infinite dilution (probe
particle porosity $\phi$). Based on the SPT2 approach, the SPT2b
approximation was proposed that reproduces the computer simulation
data with a good accuracy at small and intermediate fluid densities. However,
the  proposed approach  encounters a serious problem at high densities of a fluid.
The expressions obtained in SPT2b lead to a divergence when a packing
fraction of a fluid reaches the value equal to the probe particle porosity.
Consequently, the  prediction of thermodynamic quantities for a fluid in this
region turns out to be  wrong. The accuracy of SPT2b also decreases when fluid and
matrix particles are of comparable size  \cite{Pat16}.

In the current report we improve the SPT2 approach and propose new
approximations extending them to the case of a one-dimensional system of a
hard rod (HR) fluid in a hard rod matrix. It should be mentioned that in a
bulk one-dimensional HR fluid, the SPT theory  \cite{Helf20} reproduces the
exact result by Tonks  \cite{Tonks21}. Some analytical results for
thermodynamical properties of a HR fluid in a disordered hard rod matrix
obtained by the SPT2 approach have already been presented by us in
  \cite{Hol22}. However, these results have not been tested by a comparison
with computer simulation data. That is why to assess different approximations
in the SPT2 theory we performed grand-canonical Monte Carlo (GCMC)
simulations. We will show that for a correct description of a HR fluid in a
HR matrix it is important to introduce the third porosity $\phi^{*}$ defined
by the maximum possible value of packing fraction of a fluid confined in a
matrix.
It is observed that the new approximations lead to much better results than SPT2b.
The obtained results will be compared with the expressions for the chemical potential of a HR fluid
in a hard rod (HR) and overlapping hard
rod (OHR) matrices obtained by Reich and Schmidt using
the density functional theory~\cite{Reich23}.

It is worth noting that contrary to the three-dimensional case, the
one-dimensional model cannot be used for a description of adsorption processes, since there is
no flow in such a system due a disconnectedness of pores. Nevertheless, the considered
one-dimensional model can be useful in interpreting some processes related to
a fluid inclusion inside one-dimensional channels such as the transport
of water and ions through molecular-sized channels in  biological membranes
 \cite{Harris24}, the transportation of adsorbate molecules through zeolites
 \cite{Chen25}, the charge-carrier migration in polymers  \cite{Rich26}, etc. An
interesting example of the realization of one-dimensional model for a fluid
was demonstrated in  \cite{Wei27} by confining mesoscopic colloidal particles
in nonochannels.

The paper is organized as follows. A brief review of the SPT2 theory for a HR
fluid in HR and OHR matrices is presented in section~2. In
section~3 we formulate the new approximations within the SPT2 theory, which essentially
improve the description of thermodynamic properties of a confined fluid. Some computer simulation details will be presented in
section~4. In section~5 the comparison of different approximations of SPT2
with computer simulation results is presented and discussed. And finally we
draw some conclusions in the last section.

\section{SPT2 theory for a HR fluid in disordered HR and OHR matrices}

The basic idea of SPT is to insert an additional scaled particle of a
variable size into a fluid. This procedure is equivalent to a creation of cavity
which is free of any other fluid particles. The central point of the SPT theory
described in  \cite{Hol13,Chen15,Pat16,Reiss17,Reiss18,Lebow19,Helf20}
is a calculation of the excess chemical potential
of a scaled particle, $\mu_{\mathrm{s}}^{\mathrm{ex}}$, which is equal to a work needed to create the
corresponding cavity. In the presence of porous medium the expression for the
excess chemical potential of a small scaled particle in a one-dimensional HR fluid is as follows \cite{Hol22}:
\begin{eqnarray}
\beta\mu_{\mathrm{s}}^{\mathrm{ex}}&=&\beta\mu_{\mathrm{s}}-\ln(\rho_1\Lambda_{1})=-\ln\left[p_{0}(R_{\mathrm{s}})-\eta\left(1+\frac{R_{\mathrm{s}}}{R_{1}}\right)\right]\nonumber\\
&=&-\ln p_{0}(R_{\mathrm{s}})-\ln\left[1-\frac{\eta_{1}(1+R_{\mathrm{s}}/R_{1})}{p_{0}(R_{\mathrm{s}})}\right],
\label{hol1}
\end{eqnarray}
where $\beta=1/kT$, $k$ is the Boltzmann constant, $T$ is a temperature, $R_{\mathrm{s}}$ is
the ``radius'' (half of length) of scaled particle, $R_{1}$ is the ``radius''
of fluid particles, $\eta_{1}=2\rho_{1}R_{1}$ is the fluid packing fraction,
$\rho_{1}$ is the fluid density, $\Lambda_1$ is the fluid thermal wavelength.
The term $p_{0}(R_{\mathrm{s}})={\rm e}^{-\beta\mu_{\mathrm{s}}^{0}}$ is defined by the excess
chemical potential $\mu_{\mathrm{s}}^{0}$ of the scaled particle confined in an empty
matrix (limit of infinite dilution) and it has the meaning of the probability to
find a cavity created by this particle in the matrix in the absence of fluid
particles. The term $p_{1/0}(R_{\mathrm{s}})=1-{\eta_{1}(1+R_{\mathrm{s}}/R_{1})}{p_{0}^{-1}(R_{3})}$ is
the probability of finding a cavity created by the scaled particle in the
fluid-matrix system under the condition that the cavity is located entirely
inside a pore occupied by the scaled particle.

For a large scaled particle $(R_{\mathrm{s}}\geqslant 0)$ the excess chemical potential
$\mu_{\mathrm{s}}^{\mathrm{ex}}$ is given by a thermodynamic expression for the work needed
to create a  macroscopic cavity inside a fluid and it can be presented in
the following way
\begin{equation}
\beta\mu_{\mathrm{s}}^{\mathrm{ex}}=w(R_{\mathrm{s}})+\beta P \frac{2R_{\mathrm{s}}}{p_{0}(R_{\mathrm{s}})}\,,
\label{hol2}
\end{equation}
where $P$ is the pressure of a fluid. It is worth noting that this expression
has the same form as in a bulk case. The difference is only in the presence of a
multiplier in the last term $1/p_{0}(R_{\mathrm{s}})$. This multiplier appears due to
an excluded volume occupied by matrix particles. For the bulk case this
multiplier is absent, since $p_{0}(R_{\mathrm{s}})=1$.  However, it was also
erroneously skipped in the SPT1 theory  \cite{Hol13} leading to the
inconsistency discussed and corrected in  \cite{Pat16}. As it was already mentioned,
$p_{0}(R_{\mathrm{s}})$ is a probability of finding a cavity created by
the particle of radius $R_{\mathrm{s}}\geqslant 0$ in the matrix in the absence of fluid
particles. This probability is directly related to different types of
porosities of a matrix. First, in $R_{\mathrm{s}}=R_{1}$
\begin{equation}
p_{0}(R_{\mathrm{s}}=R_{1})={\rm e}^{-\beta\mu_{1}^{0}}=\phi
\label{hol3}
\end{equation}
one obtains the probe particle porosity, which is defined by the excess chemical potential of
fluid particles in the limit of infinite dilution $\mu_{1}^{0}$. Thus it depends on
the nature of the fluid under study. The second one is the geometrical porosity, which depends only on
the structure of a matrix and is related to the volume of the void between matrix particles,
and it is defined as $p_{0}(R_{\mathrm{s}})$ in $R_{\mathrm{s}}=0$:
\begin{equation}
p_{0}(R_{\mathrm{s}}=0)=\phi_{0}\,.
\label{hol4}
\end{equation}
The corresponding expressions for $p_{0}(R_{\mathrm{s}})$ are derived under condition $R_{\mathrm{s}}\leqslant 0$
for the case of HR matrix
\begin{equation}
p_{0}(R_{\mathrm{s}})=1-\eta_{0}\left(1+\frac{R_{\mathrm{s}}}{R_{0}}\right)
\label{hol5}
\end{equation}
and OHR matrix
\begin{equation}
p_{0}(R_{\mathrm{s}})=\exp \left[-\eta_{0}\left(1+\frac{R_{\mathrm{s}}}{R_{0}}\right)\right],
\label{hol6}
\end{equation}
respectively, where $\eta_{0}=2\rho_{0}R_{0}$,
$\rho_{0}={N_{0}}/{V}$, $N_{0}$ is the number of matrix particles, $V$ is
the ``volume'' (i.e., the length in one dimension) of a system, $R_{0}$ is the ``radius'' (half of the length)  of
a matrix particle.
Therefore, the geometrical porosity for a HR matrix has the form
\begin{equation}
\phi_{0}=p_{0}(R_{\mathrm{s}}=0)=1-\eta_{0}
\label{hol7}
\end{equation}
and for a OHR matrix it is
\begin{equation}
\phi_{0}=p_{0}(R_{\mathrm{s}}=0)={\rm e}^{-\eta_{0}}.
\label{hol8}
\end{equation}
For the probe particle porosity, $\phi$, the following expression is derived
\begin{equation}
\phi=(1-\eta_{0})\exp\left[-\frac{\eta_{0}\tau}{1-\eta_{0}}\right]
\label{hol9}
\end{equation}
in the case of HR matrix, and
\begin{equation}
\phi=\exp\left[-\eta_{0}(1+\tau)\right]
\label{hol10}
\end{equation}
in the case of OHR matrix, where $\tau=R_{1}/R_{0}$. As one can see
$\phi$ is always less than $\phi_{0}$, except the limit of $\tau=0$, in which $\phi=\phi_{0}$.

According to the ansatz of SPT, $w_{\mathrm{s}}(R_{\mathrm{s}})$ can be found from the continuity of both
expressions (\ref{hol1}) and (\ref{hol2}) for $\mu_{\mathrm{s}}^{\mathrm{ex}}$ in $R_{\mathrm{s}}=0$:
\begin{equation}
w(R_{\mathrm{s}})=-\ln\left(1-\frac{\eta_{1}}{\phi_{0}}\right).
\label{hol11}
\end{equation}
After setting $R_{\mathrm{s}}=R_{1}$, the expression (\ref{hol2}) yields the
relation between the pressure, $P$, and the excess chemical potential, $\mu_{1}^{\mathrm{ex}}$, of
a fluid in a matrix:
\begin{equation}
\beta\left(\mu_{1}^{\mathrm{ex}}-\mu_{1}^{0}\right)=-\ln\left(1-\frac{\eta_{1}}{\phi_{0}}\right)+\frac{\beta P}{\rho_{1}}\frac{\eta_{1}}{\phi}\,.
\label{hol12}
\end{equation}
Using the Gibbs-Duhem equation
\begin{equation}
\left(\frac{\partial P}{\partial\rho_{1}}\right)_{T}=\rho_{1}\left(\frac{\partial \mu_{1}}{\partial\rho_{1}}\right)_{T},
\label{hol13}
\end{equation}
an expression for compressibility is obtained
\begin{equation}
\beta\left(\frac{\partial P}{\partial\rho_{1}}\right)_{T}=\frac{1}{1-\eta_{1}/\phi}+\frac{\eta_{1}/\phi_{0}}{(1-\eta_{1}/\phi)(1-\eta_{1}/\phi_{0})}\,,
\label{hol14}
\end{equation}
which makes it possible to obtain the total chemical potential,  $\beta\mu_{1}=\ln(\rho_{1}\Lambda_{1})+\beta\mu_{1}^{\mathrm{ex}}$,
after the corresponding integration of (\ref{hol14}) over the fluid density.
Therefore, as it was shown previously in  \cite{Hol22}, the expression for the excess chemical potential
as well as for the pressure can be derived
\begin{equation}
\beta\left(\mu_{1}^{\mathrm{ex}}-\mu_{1}^{0}\right)=-\ln\left(1-\frac{\eta_{1}}{\phi}\right)-\frac{\phi}{\phi_{0}-\phi}\ln\frac{1-\eta_{1}/\phi}{1-\eta_{1}/\phi_{0}}\,,
\label{hol15}
\end{equation}
\begin{equation}
\frac{\beta P}{\rho_{1}}=-\frac{\phi_{0}}{\eta_{1}}\frac{\phi}{\phi_{0}-\phi}\ln\frac{1-\eta_{1}/\phi}{1-\eta_{1}/\phi_{0}}\,.
\label{hol16}
\end{equation}
From (\ref{hol15}) the virial (density) expansion for the chemical potential can be easily obtained in the following form
\begin{equation}
\beta\left(\mu_{1}^{\mathrm{ex}}-\mu_{1}^{0}\right)=\eta_{1}\left(\frac{1}{\phi}+\frac{1}{\phi_{0}}\right)+\frac{1}{2}\eta_{1}^{2}
\left(\frac{1}{\phi^{2}}+\frac{1}{\phi\phi_{0}}+\frac{1}{\phi_{0}^{2}}\right)+\frac{\eta_{1}^{3}}{3}
\left(\frac{1}{\phi^{3}}+\frac{1}{\phi^{2}\phi_{0}}+\frac{1}{\phi\phi_{0}^{2}}+\frac{1}{\phi_{0}^{3}}\right)+\cdots.
\label{hol17}
\end{equation}
From this expansion it is seen that the second virial coefficient is
\begin{equation}
B_{2}=2R_{1}\left(\frac{1}{\phi}+\frac{1}{\phi_{0}}\right).
\label{hol18}
\end{equation}
Later it will be shown that the second virial coefficient obtained from the SPT2 approach is in a good agreement with computer simulation results. However, the higher virial coefficients are strongly overestimated. This problem is caused by an assumption that the multiplier
$1/p_{0}(R_{\mathrm{s}})$ introduced by us in the expression (\ref{hol2}) does not depend
on the presence of a fluid. Due to this in the considered approach,
the correct value is obtained only for the second virial coefficient.

\section{Some corrections and improvements in SPT2 theory}

In order to improve the SPT2 approach we
consider two approximations proposed in  \cite{Pat16}.
In the first of them called SPT2a
the  porosity $\phi$ in (\ref{hol12}) is replaced by $\phi_{0}$. As a result one obtains
\begin{eqnarray}
\beta\left(\mu_{1}^{\mathrm{ex}}-\mu_{1}^{0}\right)^{\mathrm{SPT2a}}&=&-\ln\left(1-\frac{\eta_{1}}{\phi_{0}}\right)+\frac{\eta_{1}/\phi_{0}}{1-\eta_{1}/\phi_{0}}\,,
\label{hol3.1}
\\
\beta\frac{P^{\mathrm{SPT2a}}}{\rho_{1}}&=&\frac{1}{1-\eta_{1}/\phi_{0}}\,.
\label{hol3.2}
\end{eqnarray}
The SPT2a approximation leads to the following density expansion for the
chemical potential of a HR fluid
\begin{equation}
\beta\left(\mu_{1}^{\mathrm{ex}}-\mu_{1}^{0}\right)^{\mathrm{SPT2a}}=\frac{2\eta_{1}}{\phi_{0}}+\frac{3}{2}\frac{\eta_{1}^{2}}{\phi_{0}^{2}}+\frac{4}{3}\frac{\eta_{1}^{3}}
{\phi_{0}^{3}}+\cdots,
\label{hol3.3}
\end{equation}
from which it is seen that SPT2a does not reproduce the
second virial coefficient given in (\ref{hol18}).

The second approximation called SPT2b can be derived by replacing the logarithm
in the second term of (\ref{hol15}) with the linear term of its expansion
\begin{equation}
-\frac{\phi}{\phi_{0}-\phi}\ln\left[1-\left(\frac{\eta_{1}}{\phi}-\frac{\eta_{1}}{\phi_{0}}\right)\Big/\left(1-\eta_{1}/\phi_{0}\right)\right]\approx
\frac{\eta_{1}/\phi_{0}}{1-\eta_{1}/\phi_{0}}\,.
\label{hol3.7}
\end{equation}
Therefore, the expressions for the chemical potential and for the pressure of a HR fluid in a matrix
can be rewritten as follows:
\begin{eqnarray}
\beta\left(\mu_{1}^{\mathrm{ex}}-\mu_{1}^{0}\right)^{\mathrm{SPT2b}}&=&-\ln\left(1-\frac{\eta_{1}}{\phi}\right)+\frac{\eta_{1}/\phi_{0}}{1-\eta_{1}/\phi_{0}}\,,
\label{hol3.4}
\\
\frac{\beta P^{\mathrm{SPT2b}}}{\rho_1}&=&-\frac{\phi}{\eta_{1}}\ln\left(1-\frac{\eta_{1}}{\phi}\right)+
\frac{\phi_{0}}{\eta}\ln\left(1-\frac{\eta_{1}}{\phi_{0}}\right)+\frac{1}{1-\eta_{1}/\phi_{0}}\,.
\label{hol3.5}
\end{eqnarray}
From the SPT2b approximation the density expansion can be obtained
\begin{equation}
\beta\left(\mu_{1}^{\mathrm{ex}}-\mu_{1}\right)^{\mathrm{SPT2b}}=\eta_{1}\left[\frac{1}{\phi}+\frac{1}{\phi_{0}}\right]+
\eta_{1}^{2}\left[\frac{1}{2\phi^{2}}+\frac{1}{\phi_{0}^{2}}\right]+\eta_{1}^{3}\left[\frac{1}{3\phi^{3}}+\frac{1}{\phi_{0}^{3}}\right]+\cdots.
\label{hol3.6}
\end{equation}
As is seen, the SPT2b approximation gives the same second virial coefficient
as in the common SPT2 approach. However, the  higher virial
coefficients in SPT2b are smaller than in SPT2. Therefore, we can
expect that, similar to the three-dimensional case of a hard sphere fluid in a matrix  \cite{Pat16},
the SPT2b approximation is better than the original SPT2.

Thermodynamic quantities obtained in the SPT2 and SPT2b approximations
have divergences at high fluid densities.  From  expressions
(\ref{hol15})--(\ref{hol16}) and  (\ref{hol3.4})--(\ref{hol3.5}) it is seen that
two divergences appear at $\eta_{1}=\phi$ and $\eta_{1}=\phi_{0}$, respectively. Since
$\phi<\phi_{0}$, the first divergence ($\eta_{1}=\phi$) occurs at lower densities than the second one.
That is why overestimation for the chemical potential
in the SPT2b approximation appears very quickly when the region of high densities is reached.
To avoid this divergence, the first logarithm in the expression for chemical potential
(\ref{hol15}) (or (\ref{hol3.4})) is expanded as well
\begin{equation}
-\ln\left(1-\frac{\eta}{\phi}\right)\approx -\ln\left(1-\frac{\eta}{\phi_{0}}\right)+\frac{\eta_{1}(\phi_{0}-\phi)}{\phi_{0}\phi(1-\eta/\phi)}\label{hol3.8}\,.
\end{equation}
Therefore, a new approximation is obtained and hereafter
it is called SPT2b1 (the first correction for SPT2b). The chemical potential and pressure of a HR fluid in a matrix
in the SPT2b1 approximation are as follows:
\begin{eqnarray}
\beta\left(\mu_{1}^{\mathrm{ex}}-\mu_{1}^{0}\right)^{\mathrm{SPT2b1}}&=&-\ln\left(1-\frac{\eta_{1}}{\phi_{0}}\right)+\frac{\eta/\phi_{0}}{1-\eta/\phi_{0}}+\frac{\eta_{1}(\phi_{0}-\phi)}
{\phi_{0}\phi\left(1-\frac{\eta}{\phi_{0}}\right)}\,,
\label{hol3.9}
\\
\beta\frac{P^{\mathrm{SPT2b1}}}{\rho_{1}}&=&\frac{1}{1-\eta/\phi_{0}}\frac{\phi_{0}}{\phi}+\left(\frac{\phi_{0}}{\phi}-1\right)\frac{\phi_{0}}{\eta_{1}}
\ln\left(1-\frac{\eta_{1}}{\phi_{0}}\right).
\label{hol3.10}
\end{eqnarray}
From the expansion of chemical potential in SPT2b1
\begin{equation}
\beta\left(\mu_{1}^{\mathrm{ex}}-\mu_{1}^{0}\right)^{\mathrm{SPT2b1}}=\eta_{1}\left(\frac{1}{\phi}+\frac{1}{\phi_{0}}\right)+
\eta_{1}^{2}\left(\frac{1}{2\phi_{0}^{2}}+\frac{1}{\phi\phi_{0}}\right)+
\eta_{1}^{3}\left(\frac{1}{\phi\phi_{0}^{2}}+\frac{1}{3\phi_{0}^{3}}\right)+\cdots
\label{hol3.11}
\end{equation}
it is clearly seen that this new approximation gives the same second virial coefficient
as in SPT2 and SPT2b. The next virial coefficients are
smaller than in SPT2 and SPT2b and one can expect that the SPT2b1
approximation is better than other previous approximations.

It should be noted that the expressions for
a chemical potential of a HR fluid in HR and OHR matrices similar to (\ref{hol3.9}) were also obtained by
Reich and  Schmidt  \cite{Reich23}  using the density functional
theory. If we rewrite the expressions of Reich and Schmidt
(RS) in terms of the porosities $\phi$ and $\phi_{0}$, they will have the
form, which is more general and does not depend on the type of a matrix (HR or
OHR)
\begin{equation}
\beta\left(\mu_{1}^{\mathrm{ex}}-\mu_{1}^{0}\right)^{RS}=-\ln\left(1-\frac{\eta_{1}}{\phi_{0}}\right)+\frac{\eta_{1}/\phi_{0}}{1-\eta_{1}/\phi_{0}}+
\frac{\eta_{1}/\phi_{0}}{1-\eta_{1}/\phi_{0}}\ln\frac{\phi_{0}}{\phi}\,.
\label{hol3.12}
\end{equation}
One can see that (\ref{hol3.12}) differs from (\ref{hol3.9}) only by the coefficient in the last term
of this expression. As is seen in the limit $\phi\rightarrow\phi_{0}$, when
\begin{equation}
\ln\frac{\phi_{0}}{\phi}=\ln\left(1+\frac{\phi_{0}}{\phi}-1\right)\approx\frac{\phi_{0}}{\phi}-1,
\label{hol3.13}
\end{equation}
the expression (\ref{hol3.12}) exactly reduces to (\ref{hol3.9}), but it
does not reproduce the
second virial coefficient obtained by the SPT2 approach (\ref{hol18}).
It is seen that from (\ref{hol3.9}) and (\ref{hol3.10}) the new approximation (SPT2b1) does not have divergence in $\eta_{1}=\phi$, while the divergence in $\eta_{1}=\phi_{0}$ remains.

Also, it is worth noting that from the physical (or rather geometrical) point of view the divergence
should appear near the maximum value of fluid packing fraction,
$\eta_{1}^{\mathrm{max}}$, available for a fluid in the matrix. It is well known that for a bulk
one-dimensional system the maximum packing fraction of a fluid is equal to one, $\eta_{1}^{\mathrm{max}}=1$  \cite{Helf20,Tonks21}.
When a fluid is confined in a matrix this value should be smaller
\begin{equation}
\eta_{1}^{\mathrm{max}}=\phi^{*}\,,\qquad \phi<\phi^{*}<\phi_{0}\,.
\label{hol3.14}
\end{equation}
For a one-dimensional fluid in a HR or OHR matrix the exact expressions for $\phi^{*}$ were
obtained by Reich and  Schmidt  \cite{Reich23}. In  terms of the porosities
$\phi_{0}$ and $\phi$, the porosity $\phi^{*}$ can be presented in a general form as follows:
\begin{equation}
\phi^{*}=\frac{\phi\phi_{0}}{\phi_{0}-\phi}\ln\frac{\phi_{0}}{\phi}\,.
\label{hol3.15}
\end{equation}
This expression can be rewritten in a more compact form
\begin{equation}
v^{*}=\frac{(v-v_{0})}{\ln v/v_{0}}\,,
\label{hol3.15.1}
\end{equation}
where we introduce the quantities, which are inverse to the porosities
\begin{equation}
v_{0}=\frac{1}{\phi_{0}}\,,\qquad v=\frac{1}{\phi}\,,\qquad v^{*}=\frac{1}{\phi_{*}}\,.
\label{hol3.15.2}
\end{equation}
At the same time the quantity $v^{*}$ corresponds to the minimum of specific
volume for one fluid particle under a close packing condition. In the same
way we can interpret $v_0$ and $v$, but for the fluid densities, in which the
corresponding divergences appear for the chemical potential and pressure, i.e.
in $\eta_1=\phi_0$ and $\eta_1=\phi$. The logarithm in (\ref{hol3.15.1}),
which can be presented as
\begin{equation}
\ln\frac{v}{v_0}=-\ln\frac{\phi}{\phi_0}=\beta\left[\mu_1(R_s=R_1)-\mu_1(R_s=0)\right]
\label{hol3.15.3}
\end{equation}
is equal to the difference between the work needed to insert a scaled particle of radius $R_s=R_1$ (fluid particle)
and the work to insert a scaled particle of radius $R_s=0$ (point particle). To introduce $\phi^{*}$ into the expression for the chemical potential of a fluid (\ref{hol3.4}) the first term is modified in the following way
\begin{eqnarray}
-\ln\left(1-\frac{\eta_{1}}{\phi}\right)&=&-\ln\left(1-\frac{\eta_{1}}{\phi^{*}}\right)-\ln\left(1-\frac{\eta_{1}(\phi^{*}-\phi)}{\phi^{*}\phi(1-\eta_{1}/\phi^{*})}\right)\nonumber\\
&\approx& -\ln(1-\eta_{1}/\phi^{*})+\frac{\eta_{1}(\phi^{*}-\phi)}{\phi^{*}\phi(1-\eta_{1}/\phi^{*})}\,.
\label{hol3.16}
\end{eqnarray}
Such a procedure leads to a new approximation which hereafter is called SPT2b2 (the second correction to SPT2b).
Therefore, the corresponding expressions for the excess chemical potential and pressure are as follows:
\begin{eqnarray}
\beta\left(\mu_{1}^{\mathrm{ex}}-\mu_{1}^{0}\right)^{\mathrm{SPT2b2}}&=&-\ln\left(1-\frac{\eta_{1}}{\phi^{*}}\right)+\frac{\eta_{1}/\phi_{0}}{1-\eta_{1}/\phi_{0}}
+\frac{\eta(\phi^{*}-\phi)}{\phi^{*}\phi(1-\eta_{1}/\phi^{*})}\,,
\label{hol3.17}
\\
\frac{\beta P^{\mathrm{SPT2b2}}}{\rho_{1}}&=&-\frac{\phi^{*}}{\eta_{1}}\ln\left(1-\frac{\eta_{1}}{\phi^{*}}\right)+\frac{\phi_{0}}{\eta_{1}}\ln \left(1-\frac{\eta_{1}}{\phi_{0}}\right)
+\frac{1}{1-\eta_{1}/\phi_{0}}\nonumber\\
&&{}+\frac{\phi^{*}-\phi}{\phi}%
\left[\ln\left(1-\frac{\eta_{1}}{\phi^{*}}\right)+\frac{\eta_{1}/\phi^{*}}{1-\eta_{1}/\phi^{*}}\right].
\label{hol3.18}
\end{eqnarray}
The density expansion of the chemical potential of a fluid in the SPT2b2 approximation can be derived as
\begin{eqnarray}
\beta\left(\mu_{1}^{\mathrm{ex}}-\mu_{1}^{0}\right)^{\mathrm{SPT2b2}}=\eta\left(\frac{1}{\phi}+\frac{1}{\phi_{0}}\right)+
\eta_{1}^{2}\left[\frac{1}{2\phi^{*2}}+\frac{1}{\phi_{0}^{2}}+\frac{\phi^{*}-\phi}{\phi^{*2}\phi}\right]%
+\eta_{1}^{3}\left(\frac{1}{3\phi^{*3}}+\frac{1}{\phi_{0}^{3}}+\frac{\phi^{*}-\phi}{\phi^{*3}\phi}\right)+\cdots.
\label{hol3.19}
\end{eqnarray}

The third correction for SPT2b is obtained by expanding
the first logarithm in (\ref{hol3.17}) and it is referred to in this paper as the SPT2b3 approximation.
This procedure is similar to that in (\ref{hol3.16}), thus the excess chemical potential
and pressure of a fluid in a matrix in the SPT2b3 approximation can be easily derived as follows:
\begin{eqnarray}
\beta\left(\mu_{1}^{\mathrm{ex}}-\mu_{1}^{0}\right)^{\mathrm{SPT2b3}}&=&-\ln\left(1-\frac{\eta_{1}}{{\phi_{0}}}\right)+\frac{\eta_{1}/\phi^{*}}{1-\eta_{1}/\phi_{0}}
+\frac{\eta_{1}(\phi^{*}-\phi)}{\phi^{*}\phi(1-\eta_{1}/\phi^{*})}\,,
\label{hol3.20}
\\
\frac{\beta P^{\mathrm{SPT2b3}}}{\rho_{1}}&=&\frac{1}{1-\eta_{1}/\phi_{0}}+\frac{\phi^{*}-\phi}{\phi}\left[\ln\left(1-\frac{\eta_{1}}{\phi^{*}}\right)+
\frac{\eta_{1}/\phi^{*}}{1-\eta_{1}/\phi^{*}}\right]\nonumber\\
&&{}+\frac{\phi_{0}-\phi^{*}}{\phi^{*}}\left[\ln\left(1-\frac{\eta_{1}}{\phi_{0}}\right)+\frac{\eta_{1}/\phi_{0}}{1-\eta_{1}/\phi_{0}}\right].
\label{hol3.21}
\end{eqnarray}
The SPT2b3 approximation yields the corresponding density expansion of the  chemical potential
\begin{eqnarray}
\beta\left(\mu_{1}^{\mathrm{ex}}-\mu_{1}^{0}\right)^{\mathrm{SPT2b3}}=\eta_{1}\left(\frac{1}{\phi}+\frac{1}{\phi_{0}}\right)+\eta_{1}^{2}\left[\frac{1}{2\phi_{0}^{2}}+\frac{1}
{\phi^{*}\phi_{0}}+\frac{\phi^{*}-\phi}{\phi^{*2}\phi}\right]%
+\eta_{1}^{3}\left[\frac{1}{3\phi_{0}^{3}}+\frac{1}
{\phi^{2}_{0}\phi^{*}}+\frac{\phi^{*}-\phi}{\phi^{*3}\phi}\right]+\cdots.
\label{hol3.22}
\end{eqnarray}
From (\ref{hol3.19}) and (\ref{hol3.22}) it is seen that the both approximations
SPT2b2 and SPT2b3 reproduce the second virial coefficient presented in
(\ref{hol18}). The principal difference of the last two approximations
comparatively with the other approximations considered in this study is that
the SPT2b2 and SPT2b3 approximations have the divergence in the correct value
of packing fraction $\eta_{1}=\phi^{*}$, which corresponds to the maximum
value of packing fraction of a HR fluid in a matrix. Therefore, one can expect
that these approximations provide a correct description of a HR fluid in a
matrix at high fluid densities. The accuracy of all the considered approximations
will be tested by a comparison with computer simulation
results.

 \section{Computer simulations details}
The grand canonical Monte Carlo simulations  \cite{Pats24} were performed
within this study in order to verify the accuracy of the theory proposed. To
this end, a one-dimensional system of hard rods was considered. The system
consisted of two components. The first component is described by unmovable hard
rod particles of a matrix. The second one is described by hard rod particles of a fluid
that could move in a space not occupied by the matrix particles. A hard rod
interaction between particles in the system is defined by radii $R_0$ and
$R_1$, which are half sizes of matrix and fluid particles respectively. We
considered two types of a matrix with different structures. A structure of
the first matrix is modeled as an equilibrated one-dimensional HR fluid, thus
it is referred to as a HR matrix. The second one presents HR particles which are randomly
distributed in space and can overlap, and this model is referred to as
an overlapping hard rod (OHR) matrix.

To compose configurations of a HR matrix, the preliminary simulations were carried out  in canonical ensemble.
For this purpose, a number of matrix particles equal to $N_0=10000$ were taken and after equilibration had been reached
the coordinates of them were stored to use them in the main simulations for a confined fluid.
In the case of a OHR matrix, the coordinates of matrix particles were randomly set.

The different sizes of matrix particles were considered in simulations, starting with the limit
case of zero size and finishing with the large enough matrix particle sizes
($R_0=0.0\sigma$; $R_0=0.5\sigma$; $R_0=1.5\sigma$; $R_0=2.5\sigma$). It should be noted
that all sizes and lengths in our study are presented in units of $\sigma$, which is equal to a diameter of fluid particles.
Thus, the radius of fluid particles is $R_1=0.5\sigma$. Another important parameter for a matrix
is the porosity $\phi$, which was fixed in our study and was equal to $0.35$. Therefore, depending
on the size of matrix particles, the density of matrix particles was adjusted to make $\phi$ equal to this value
using the expressions (\ref{hol9}) and (\ref{hol10}) for HR and OHR matrices, respectively.
Since, a number of matrix particles, $N_0$, were fixed, the volume of the box depended on the matrix density:
\begin{align}
V=L=\frac{N_0}{\rho_0}\,,
\label{hol4.1}
\end{align}
where $\rho_{0}=\eta_{0}/2R_{0}$. A number of fluid particles varied
depending on the chemical potential of a fluid up to $N_{1}\sim15000$. During each
simulation run, equilibration and production procedures were performed. The number
of steps for equilibration equal to $600000$, and $200000$ steps for the production  appeared to be sufficient to get reliable results for the fluid density
at the given chemical potentials. It should be also noted that the results show
a remarkable dependence on a matrix configuration. That is why in such a case,
 different configurations of a matrix should be used to take an average over
them. To this end, we used $8$ matrix configurations for each set of matrix
parameters and the chemical potential to get a satisfactory statistical error
for the fluid density, which was less than $0.5\%$. Therefore, from the simulations we obtained
numerical relations between the fluid density and the chemical potential, which were
compared with the results calculated from the SPT2 approach. The
corresponding analysis is made in the next section.

\section{Results and discussions}

Above we presented the analytical expressions for the chemical potential and the pressure,
which were obtained in different approximations for a one-dimensional hard rod (HR) fluid confined in a disordered matrix. These expressions include three types of porosities, i.e. the geometrical porosity $\phi_{0}$,
the specific probe particle porosity $\phi$ and the porosity $\phi^{*}$
defined by the maximum packing fraction of a fluid in a given matrix.
All these porosities $\phi_{0}$, $\phi$ and $\phi^{*}$ have exact expressions
(\ref{hol7})--(\ref{hol10}) and (\ref{hol3.15}).

Before testing the accuracies of the proposed approximations we consider two limits. One of them
$\tau=R_{1}/R_{0}\rightarrow0$ was discussed previously in  \cite{Pat16} for a hard
sphere fluid in a matrix. In this limit, a size of matrix particles
tends to infinity, $R_{0}\rightarrow\infty$, while the packing fraction of matrix particles $\eta_{0}$ is fixed.
In this limit one can see that $\phi_{0}=\phi=\phi^{*}$ and the considered
approximations lead to the same result, which corresponds to a bulk fluid with
a density $\hat{\eta}_{1}=\eta_{1}/\phi_{0}$. In the second limit
$\tau\rightarrow\infty$ and a size of matrix particles tends to zero, $R_{0}\rightarrow0$ ($\eta_0\rightarrow0$),
while $\eta_{0}\tau=2\rho_{0}R_{1}$ is kept constant. This limit corresponds
to the matrix of point particles and in this case the difference between
HR and OHR matrices disappears. The expressions of porosities for a point matrix are as
follows:
\begin{equation}
\phi_{0}=1, \qquad \phi={\rm e}^{-\eta_{0}\tau}\,,\qquad \phi^{*}=\frac{\tau\eta_{0}}{{\rm e}^{\tau\eta_{0}}-1}\,.
\label{hol5.1}
\end{equation}

Now we proceed to assess the accuracy of different approximations of SPT2 at
different values of $\tau$ and fluid densities. For convenience, we
fix the probe particle porosity, $\phi$, and for all cases considered in this study
we set $\phi=0.35$. For $\tau$ we consider three cases $\tau=1$, $\tau=1/3$ and
$\tau=1/5$. We start with a comparison of the approximations SPT2, SPT2a and
SPT2b with the results of grand canonical Monte-Carlo simulations obtained in this study (figure~1).
Similar to the three-dimensional case  \cite{Pat16} SPT2a
underestimates and SPT2 overestimates the values of chemical potential in comparison with GCMC.
The SPT2b approximation improves the results for low and intermediate fluid densities.
With a decreasing $\tau$ the improvement becomes more essential. Since SPT2 and SPT2b have the divergence
at $\eta_{1}=\phi$, the description is not correct at higher densities. This drawback
should be eliminated by the new approximations SPT2b1, SPT2b2 and SPT2b3.
In figure~2 a comparison of these approximations with the SPT2b approximation and with the expressions by Reich and Schmidt \cite{Reich23} are presented (in the figure denoted as RS results).
As one can see SPT2b1 essentially improves the  results at intermediate densities, but with
a remarkable underestimation at high fluid densities. It is observed that the RS results are rather close
to that obtained in SPT2b1, but it yields a little bit lower values of the chemical potential.
With a decrease of $\tau$, the difference between these two approximations reduces.
However, due to the divergence in $\eta_{1}=\phi_{0}$ appearing in these two approximations,
the chemical potential can be finite at $\eta_{1}>\phi^{*}$ ($\phi^{*}<\phi_{0}$)
\begin{figure}[!h]
\begin{center}
\begin{tabular}{c c}
  \includegraphics[width=7.2cm]{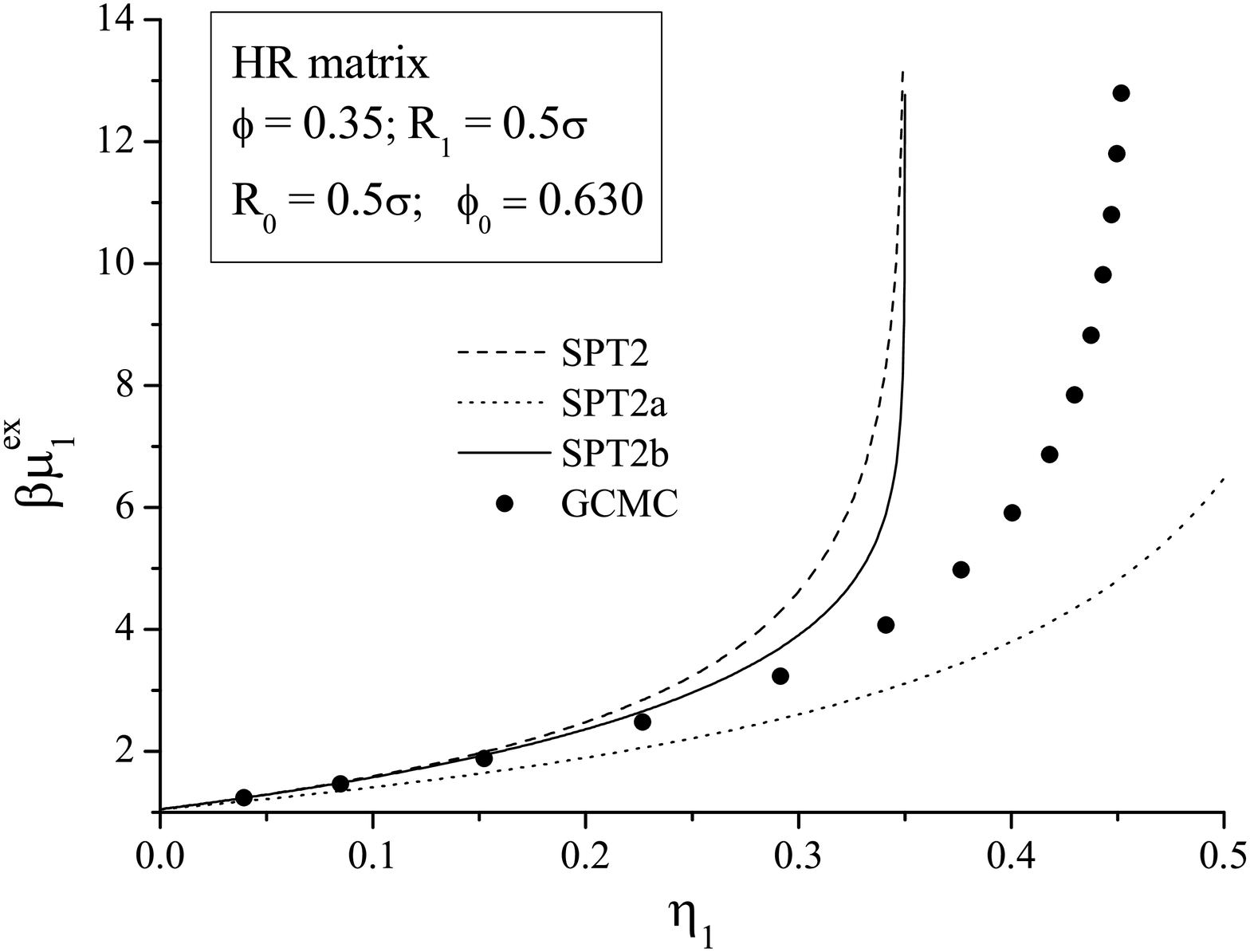}
  \includegraphics[width=7.2cm]{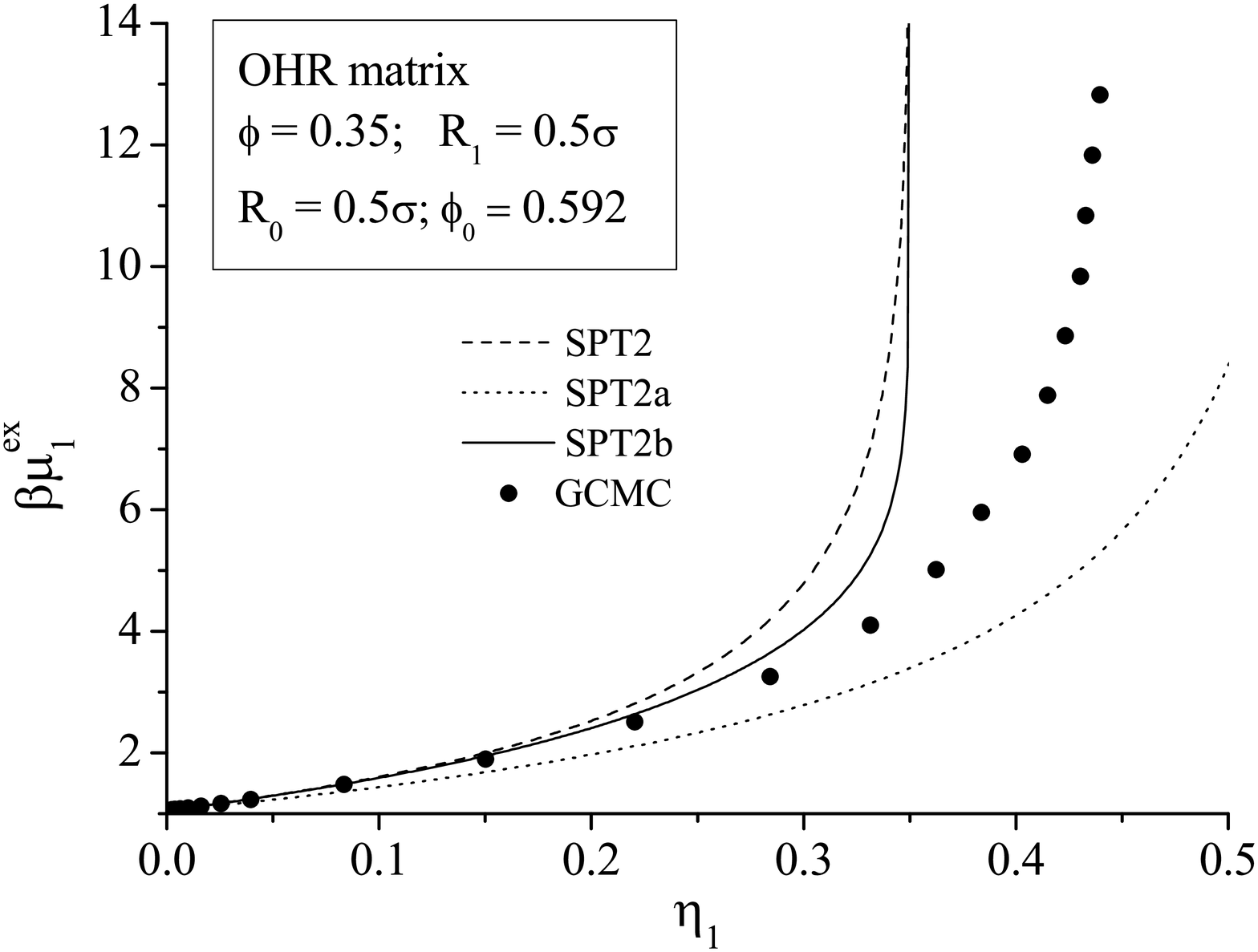}\\
  \includegraphics[width=7.2cm]{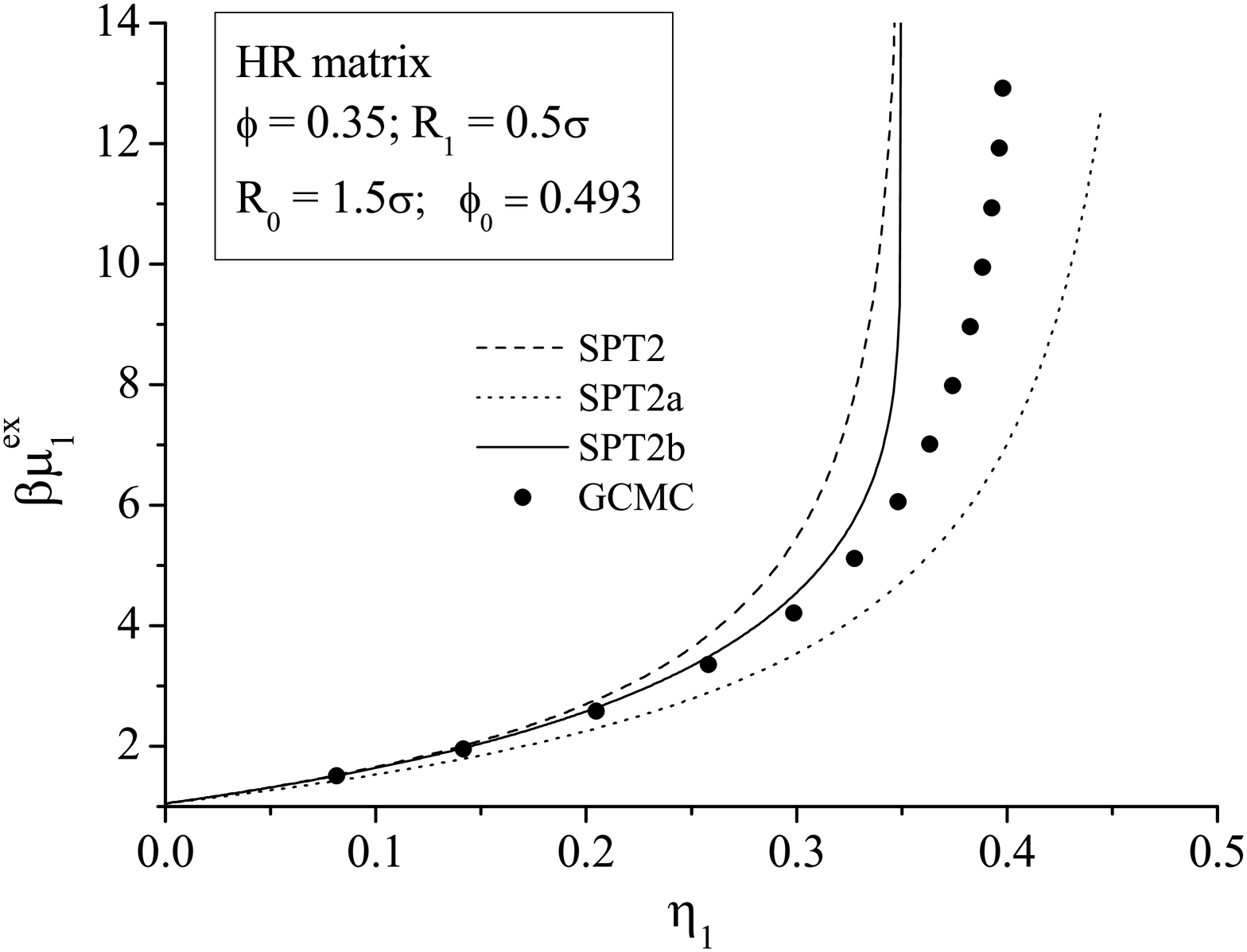}
  \includegraphics[width=7.2cm]{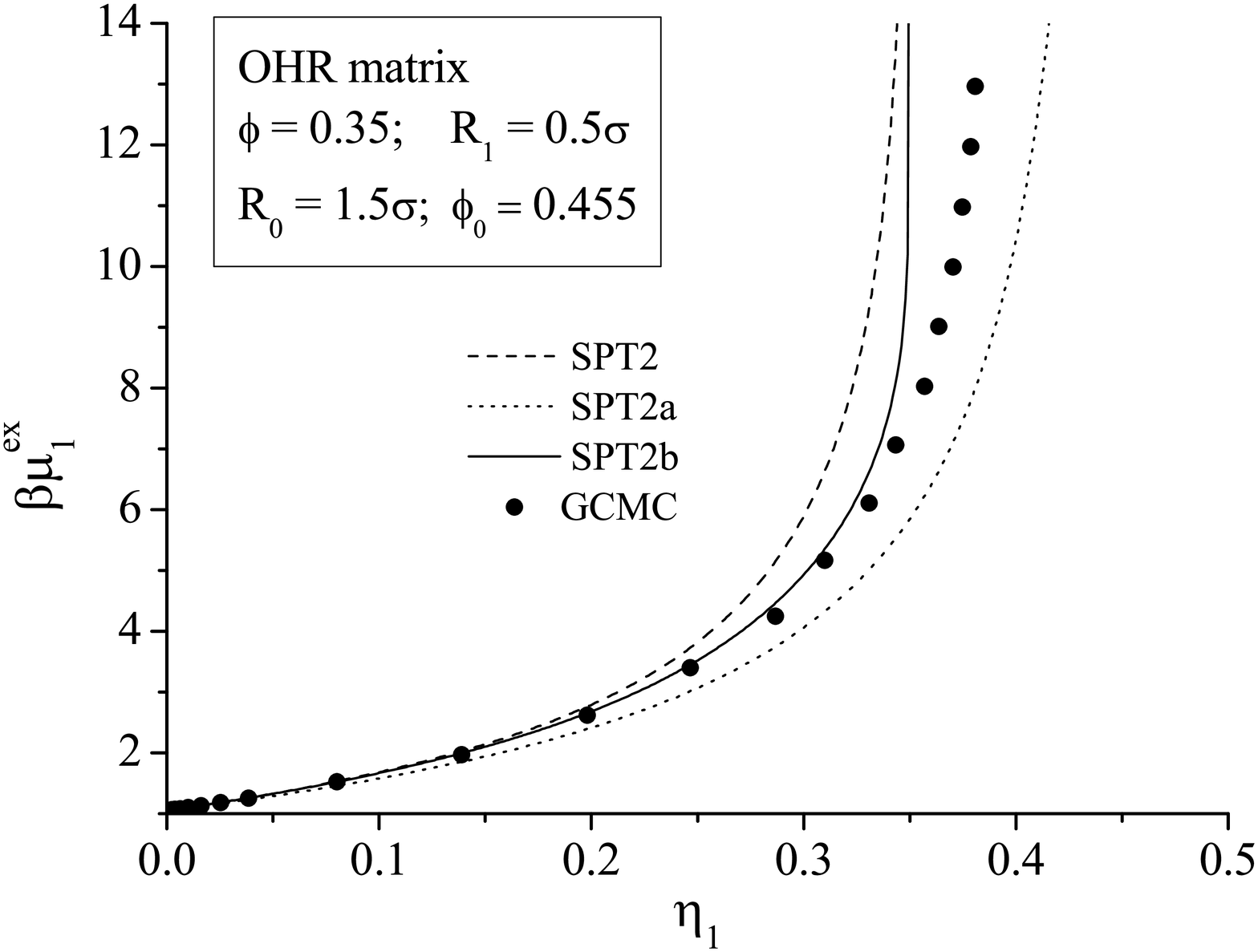} \\
  \includegraphics[width=7.2cm]{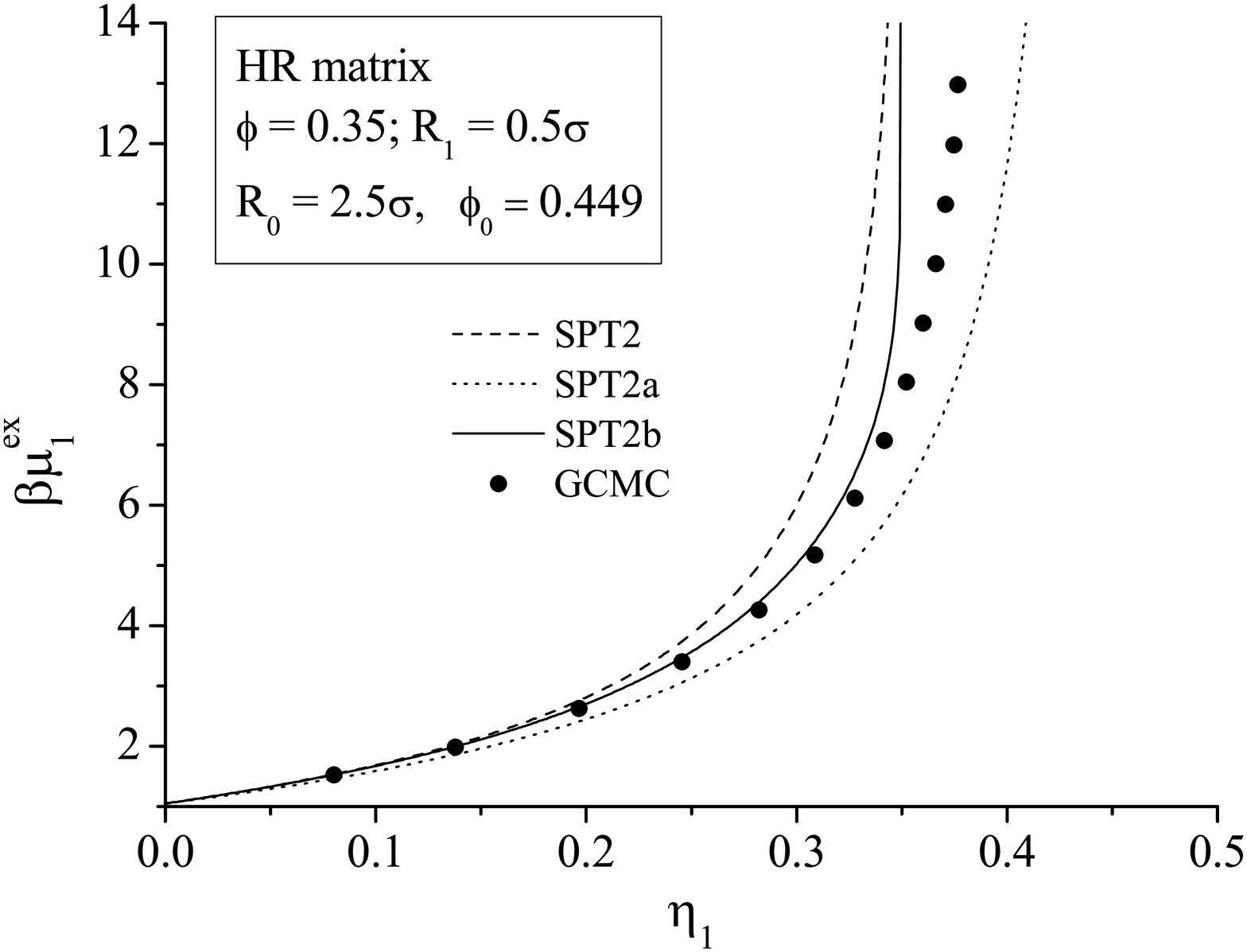}
  \includegraphics[width=7.2cm]{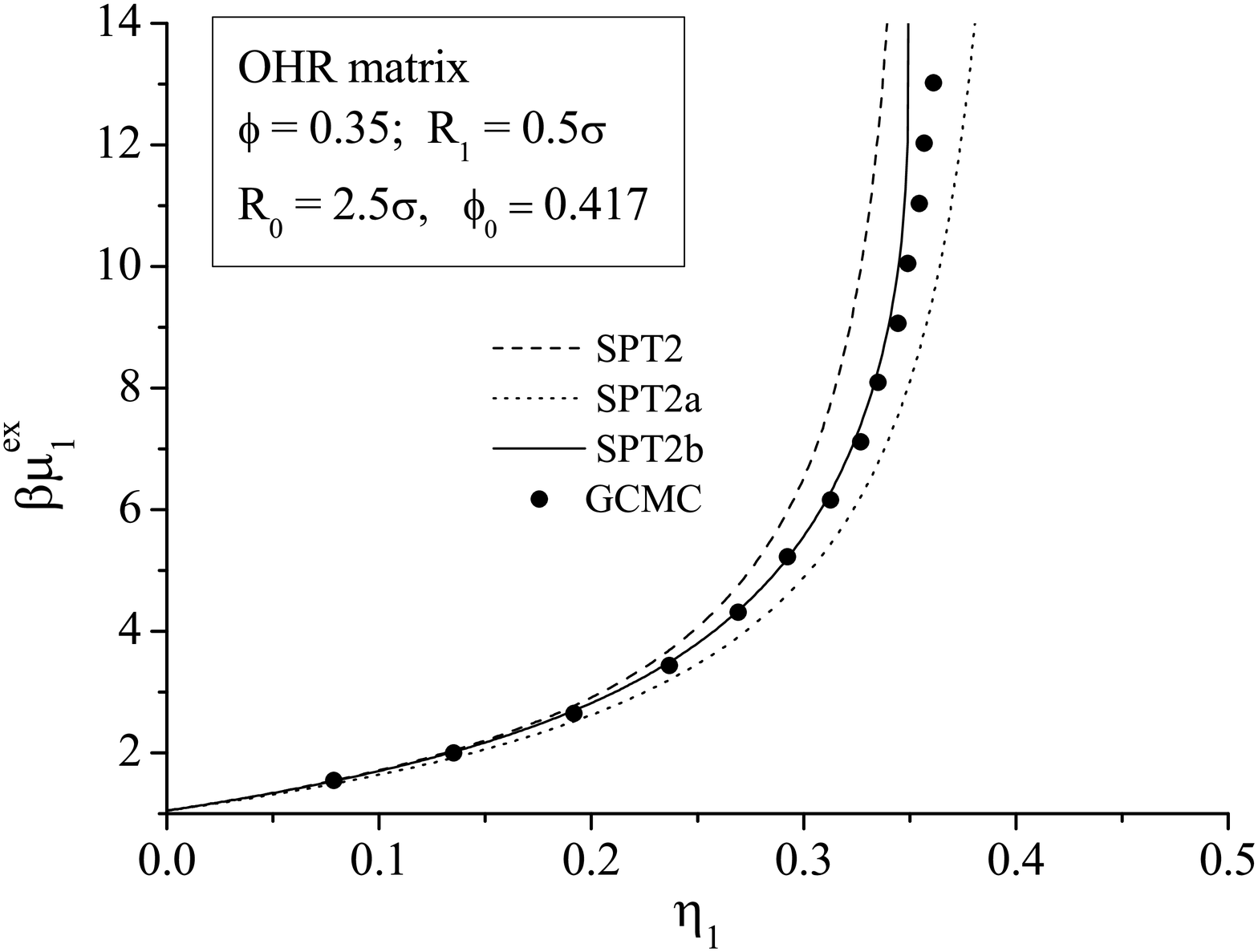} \\
\end{tabular}
\caption{The excess chemical potential of a one-dimensional HR fluid in a disordered HR (left panel)
or OHR (right panel) matrix. A comparison of SPT2, SPT2a and SPT2b approximations (lines) with the GCMC simulation
results (symbols).}
\label{Fig1}
\end{center}
\end{figure}
and this is not correct since
the packing fraction of a fluid cannot reach the values that are higher than the maximum packing
fraction available for a fluid in a matrix, $\phi^{*}$. The  best description is provided by the SPT2b2 and
SPT2b3 approximations, which have a correct divergence at $\eta_{1}=\phi^{*}$, while
SPT2b3 is slightly better than SPT2b2.

\begin{figure}[!h]
\begin{center}
\begin{tabular}{c c}
  \includegraphics[width=7.2cm]{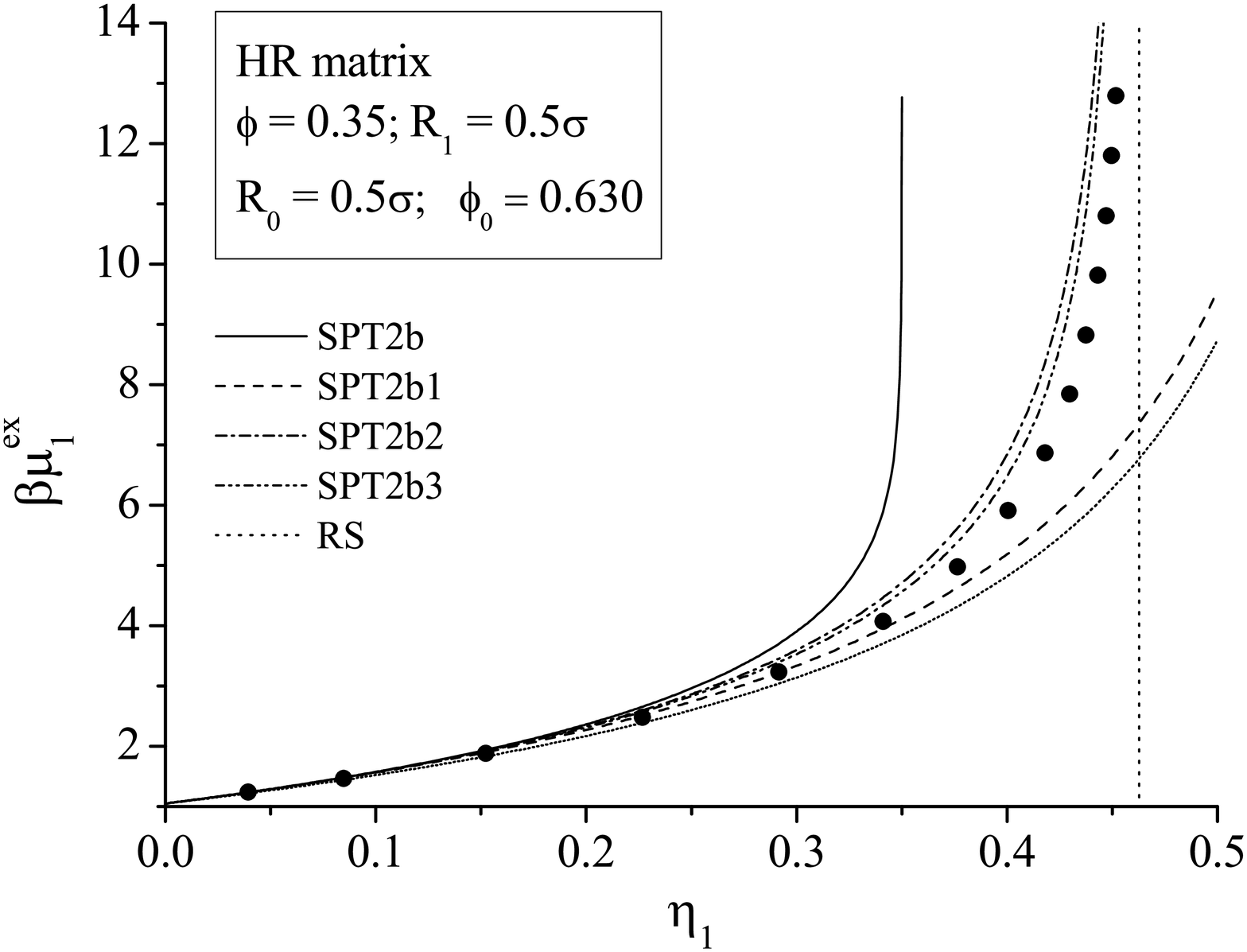}
  \includegraphics[width=7.2cm]{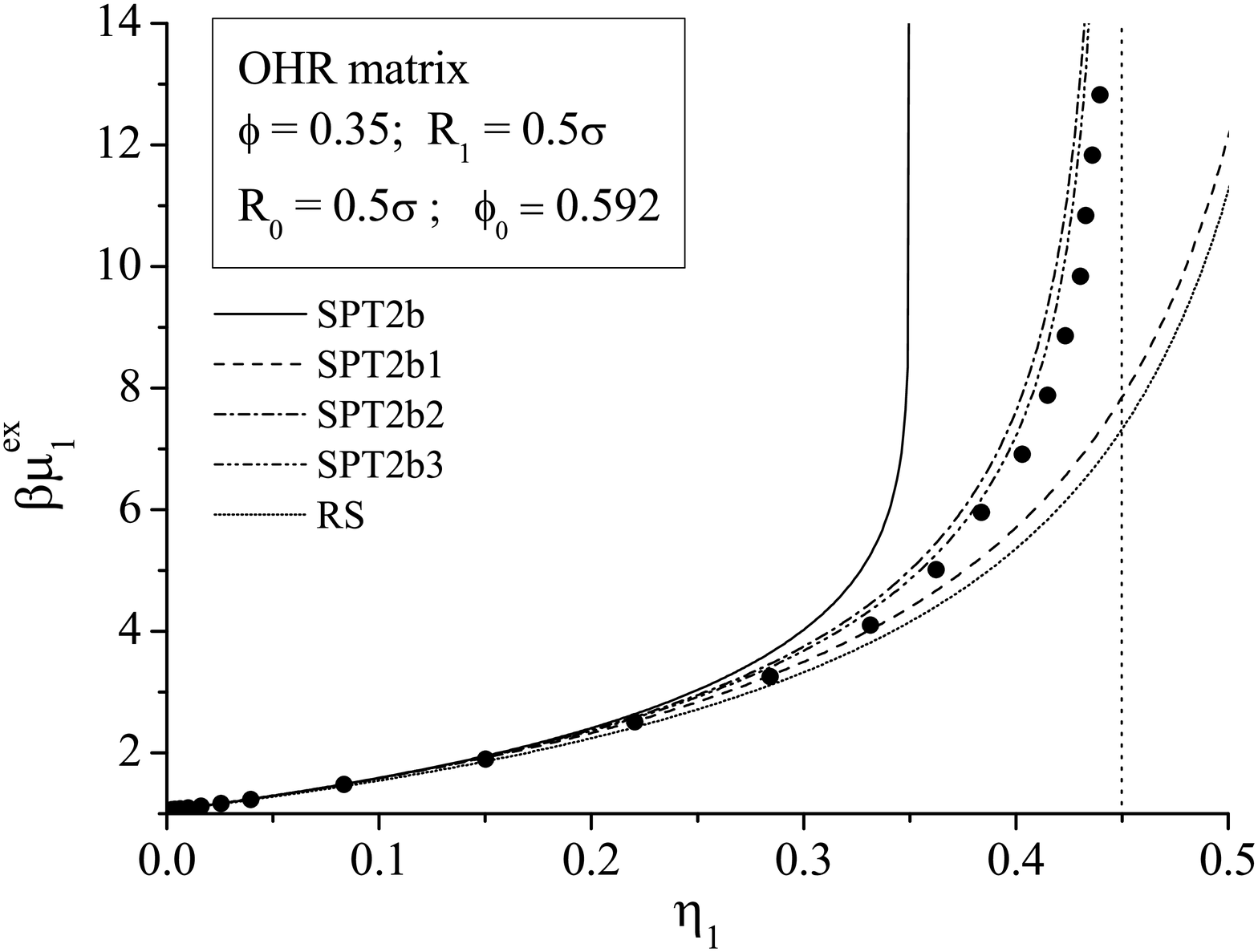}\\
  \includegraphics[width=7.2cm]{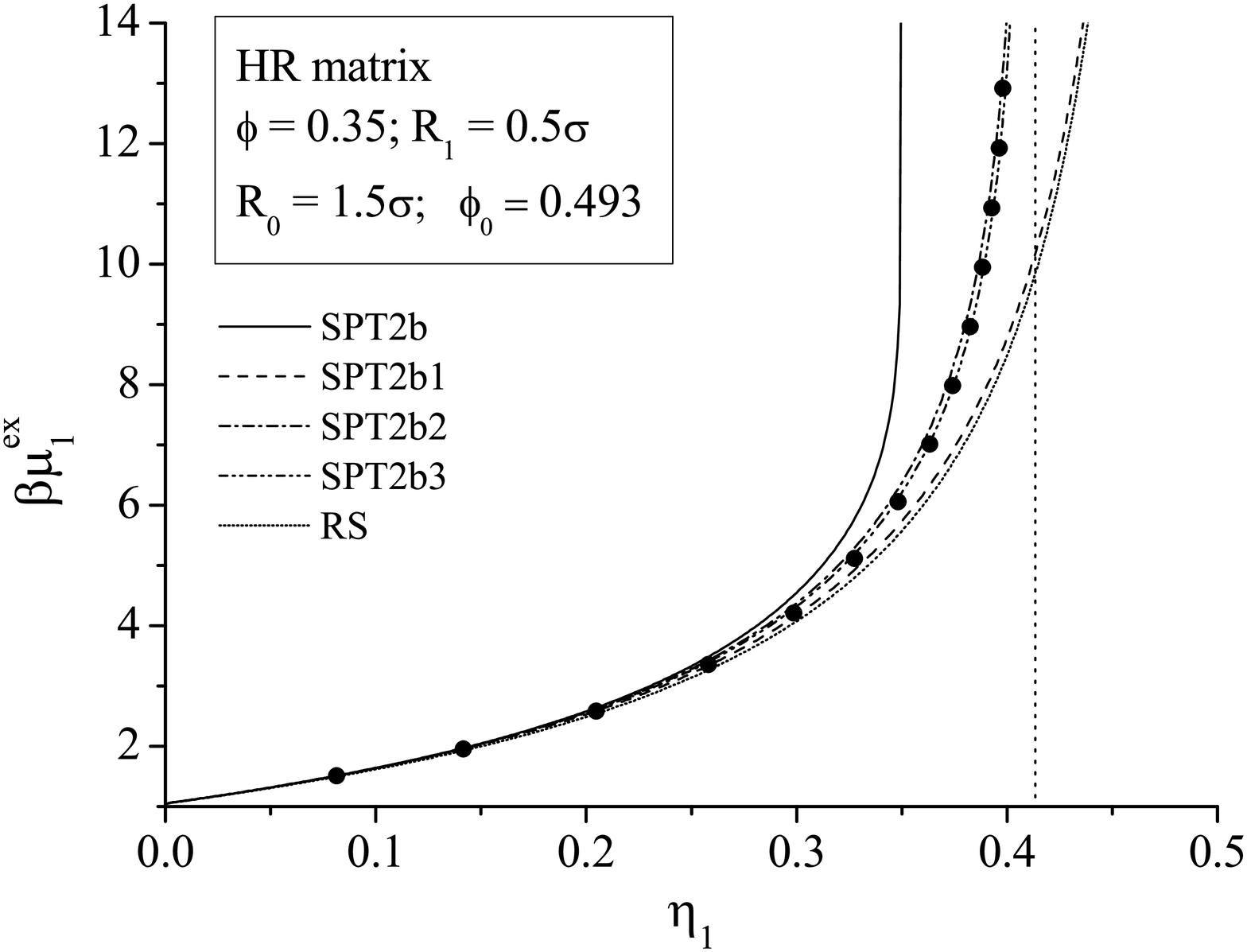}
  \includegraphics[width=7.2cm]{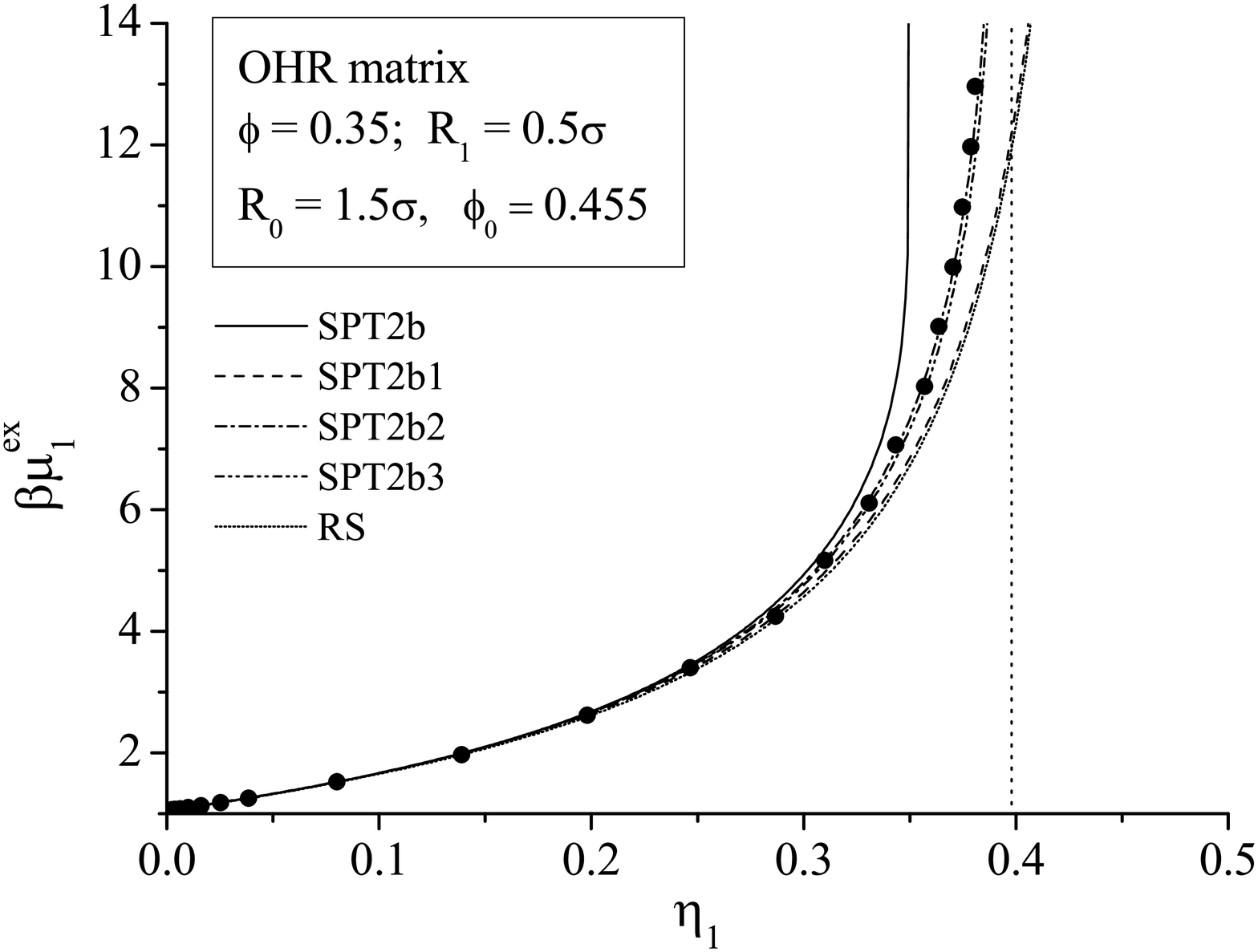} \\
  \includegraphics[width=7.2cm]{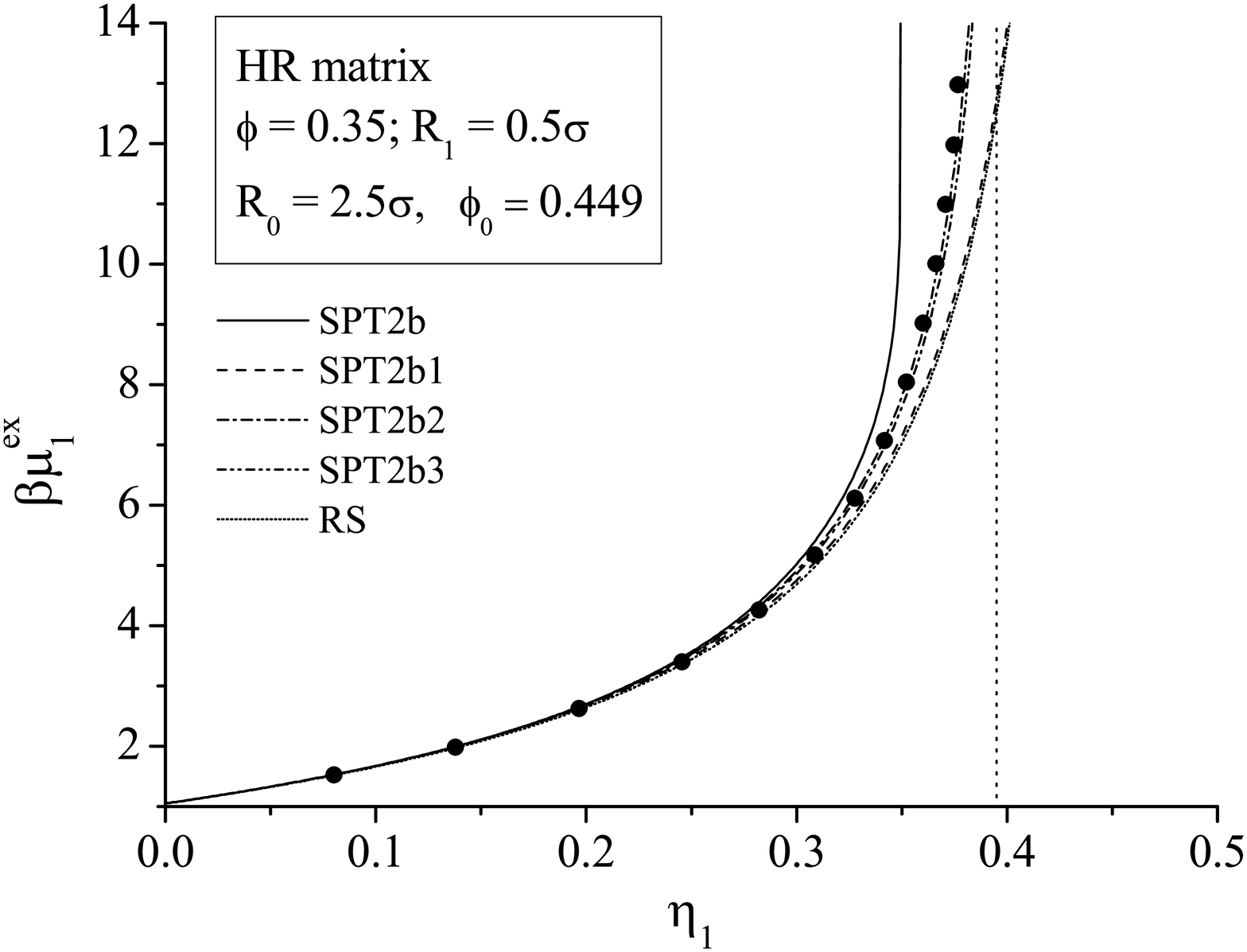}
  \includegraphics[width=7.2cm]{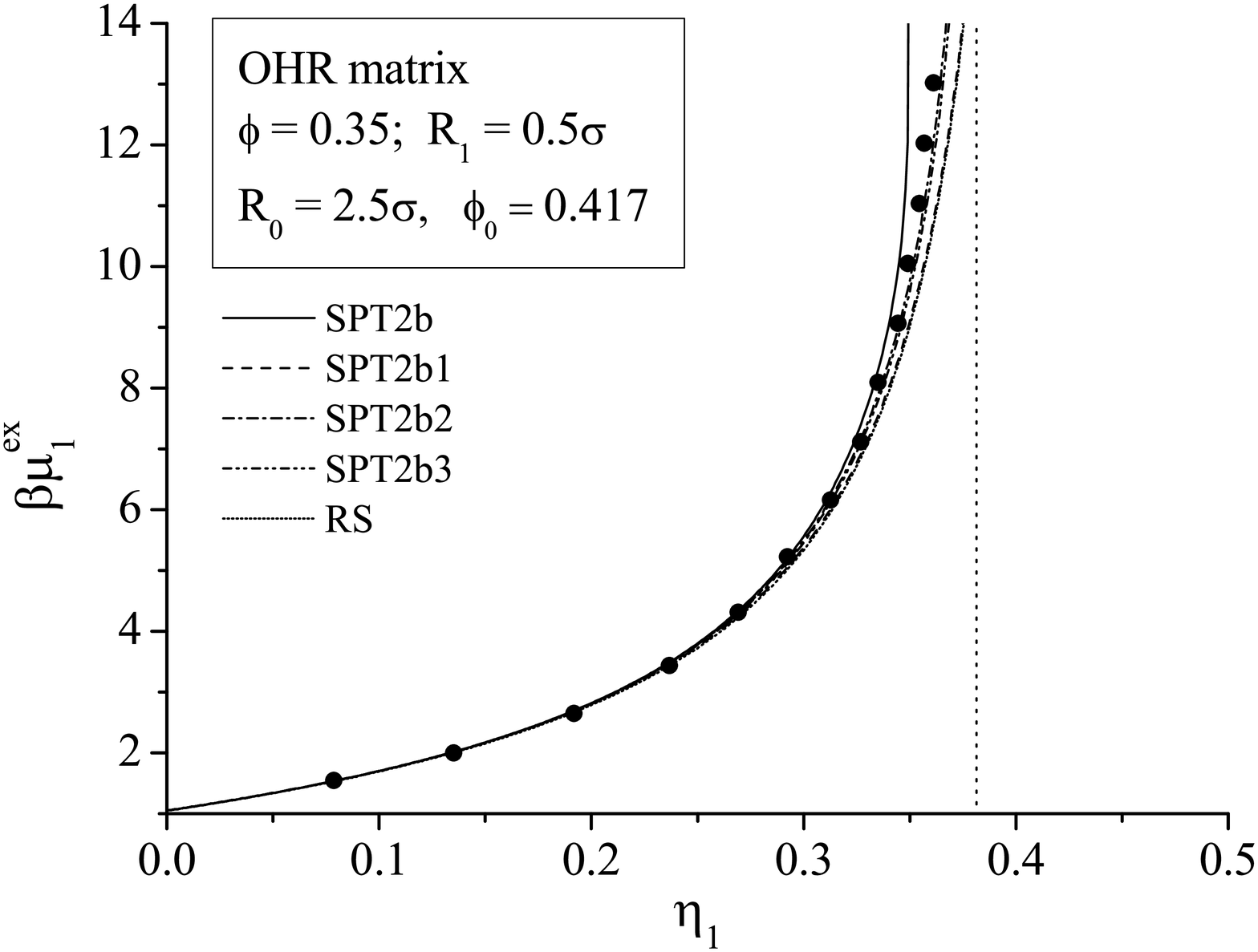} \\
\end{tabular}
\caption{The excess chemical potential of a one-dimensional HR fluid in a disordered HR (left panel)
or OHR (right panel) matrix. A comparison of different approximations (lines) with the GCMC simulation
results (symbols). The vertical dotted line corresponds to the value of maximum packing fraction of
a one-dimensional fluid confined in a HR or OHR matrix, $\phi^{*}$.}
\label{Fig2}
\end{center}
\end{figure}
The results for a one-dimensional HR fluid in a point matrix $(\tau\rightarrow\infty)$ are
presented in figure~3. As one can see, in this case, SPT2b is correct only up
to the density $\eta_{1}=0.2$. SPT2b1 provides a correct description up to
$\eta_{1}=0.45$. In this case, the difference between SPT2b1 and
RS approximations is more pronounced and SPT2b1 shows a better agreement with the simulations, while SPT2b2
and SPT2b3 provides a the correct description up to $\eta_{1}=0.3$. They
also provide a qualitatively correct description of the chemical potential at
high fluid densities.

\begin{figure}[!h]
\begin{center}
\begin{tabular}{c}
  \includegraphics[width=7.2cm]{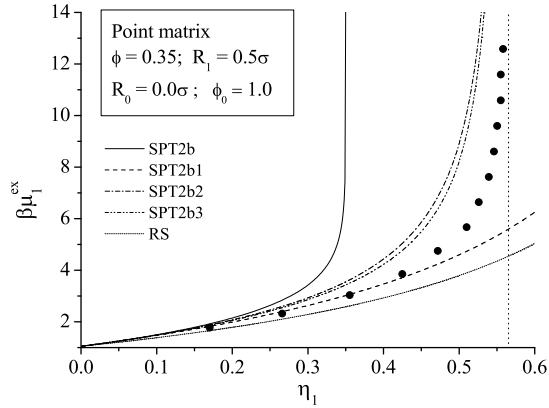}
\end{tabular}
\caption{The excess chemical potential of a one-dimensional HR fluid in a disordered point matrix.
A comparison of different approximations (lines) with the GCMC simulation
results (symbols). The vertical dotted line corresponds to the value of maximum packing fraction of
a one-dimensional fluid confined in a point matrix, $\phi^{*}$.}
\label{Fig3}
\end{center}
\end{figure}
\begin{figure}[!h]
\begin{center}
\begin{tabular}{c}
  \includegraphics[width=7.2cm]{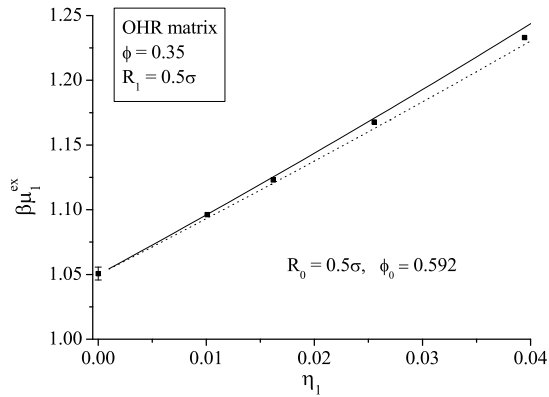}
\end{tabular}
\caption{The excess chemical potential of a one-dimensional HR fluid in a OHR matrix
at very low densities. Solid line denotes the SPT2 approach and dotted line corresponds to
the result obtained by the RS formula  \cite{Reich23}.}
\label{Fig4}
\end{center}
\end{figure}
Finally, we consider the description of chemical potential at small densities,
where it is sufficient to take into account only the second virial coefficient.
In sections~2 and 3 it is shown that for all of the considered approximations, except the SPT2a and RS,
the second virial coefficient has the same expression (\ref{hol18}).
In figure~4 the dependence of the chemical potential of a HR fluid in a OHR matrix on a fluid density is shown at
the small densities up to $\eta=0.04$. As one can see, in this region
the chemical potential is correctly described by the second
virial coefficient in the form (\ref{hol18}). The second
virial coefficient obtained from the RS approach leads to a slight underestimation of the chemical potential
in the region of the considered fluid densities.

\section{Conclusions}

In this  paper the recently proposed scaled particle theory SPT2
 \cite{Pat16} for  the description of hard-sphere fluid in disordered porous
media is applied to a one-dimensional hard rod fluid in HR and OHR matrices. The  obtained
analytical expressions for the  chemical potential and pressure of a HR
fluid in a disordered matrix explicitly include two types of porosity. One of
them is geometrical porosity, $\phi_{0}$, describing the free void
in a porous media, and  the  second one is the probe particle porosity,
$\phi$, described by an adjustable volume for a fluid in a porous medium and
is defined by the chemical potential of a fluid in the limit of infinite dilution. It is
shown that the SPT2 approach correctly reproduces the second virial coefficient, but
overestimates the higher virial coefficients. In order to improve
the SPT2 theory we have applied two approximations -- SPT2a and SPT2b.
To assess these approximations, the extensive GCMC simulations were
carried out. A detailed comparison of the  results of the
approximations with the simulations shows that all of them have got a
good accuracy at small and intermediate fluid densities.
However, SPT2a leads to a remarkable underestimation of the chemical potential while SPT2b causes overestimation
at high densities.

The principal defect of the SPT2 approach is
related to the divergence of thermodynamical properties at the densities
corresponding to the probe particle porosity $\eta_{1}=\phi$.
In order to avoid this defect three new approximations were proposed. In one
of them called SPT2b1,  thermodynamical properties have a divergence
only at the density corresponding to the geometrical porosity
$\eta_{1}=\phi_{0}$. This approximation provides a very accurate description
at rather high densities for different parameters of porous media
including the matrices formed by disordered point particles. To
describe thermodynamical properties in the region of very high fluid
densities we have formulated two new approximations SPT2b2 and SPT2b3, which explicitly
include the third porosity $\phi^{*}$ defined by the maximum value
of packing fraction of a fluid in a matrix. These approximations have a
divergence at $\eta_{1}=\phi^{*}$, which provides a correct description
of thermodynamical properties at high densities.

It is worth noting that the theory proposed and validated in this report can
be easily extended to the cases of two- and three- dimensional systems.
The application of the developed theory in combination with the new approximations for a three-dimensional hard sphere fluid in  hard sphere matrices will be described elsewhere.

\section*{Acknowledgements}
M.H. and T.P. thank the National Academy of Sciences of Ukraine for the
support of this work (the joint NASU--RFFR Program).
All simulations were performed at the computing cluster of the Institute
for Condensed Matter Physics (Lviv).

\ukrainianpart

\title{Одновимірний твердострижневий плин у невпорядкованому пористому
середовищі: \\теорія масштабної частинки}
\author{М.~Головко\refaddr{label1}, Т.~Пацаган\refaddr{label1}, В.~Донг\refaddr{label2}}
\addresses{
\addr{label1} Інститут фізики конденсованих систем НАН України, Україна, 79011 Львів,
вул.~Свєнціцького~1.
\addr{label2} Вища нормальна школа Ліону, Лабораторія хімії, Франція
}
%
%
%

\makeukrtitle

\begin{abstract}
\tolerance=3000%
Застосовано теорію масштабної частинки  до опису термодинамічних властивостей плину в матриці в рамках моделі твердих стержнів. З цією метою розвинуто теоретичний підхід, відомий як SPT2. Отримано аналітичні вирази для хімічного потенціалу і тиску плину в матриці двох типів: тверді стержні та тверді стержні, що перекриваються. Запропоновано низку нових наближень в рамках SPT2. Показано, що, крім відомих геометричної пористості та пористості пробної частинки, важливим є ще інший тип пористості, який визначається максимально можливою упаковкою плину, доступною в пористому середовищі. Щоб перевірити теорію SPT2 та оцінити точність запропонованих наближень, проведено комп'ютерне моделювання методом Монте-Карло у великому канонічному ансамблі. Зауважено, що теоретичний опис, представлений у даному дослідженні, суттєво покращує результати аж до найвищих густин плину.
\keywords плини в просторових обмеженнях, пористі матеріали, теорія масштабної частинки,
твердострижневий плин, термодинамічні властивості, комп'ютерне моделювання

\end{abstract}

\end{document}